# Landing the uniformly accelerating observers

Bernhard Rothenstein[1], Stefan Popescu [2] and Ronald P. Gruber [3]
1) Politehnica University of Timisoara, Physics Department,
Timisoara, Romania brothenstein@gmail.com
2) Siemens AG, Erlangen, Germany stefan.popescu@siemens.com
3) SUMC, Stanford University, Stanford, USA

***Abstract.*** *Observers of the uniformly accelerating observers or the observers who make up the system of uniformly accelerating observers reach the same velocity V at different times $t_i$ which depends on V and on theirs acceleration $g_i$. Considering a platform that moves with constant velocity V, the observers can land on it smoothly. Their ages and locations in the inertial reference frame attached to the platform are reckoned and compared.*

## 1. Introduction

For the purpose of the analysis below we will use the same scenario that we considered in a previous paper[1]: point-like observers equipped with identical clocks are initially in state of rest being located at the origin O of the K(XOY) inertial reference frame. When all their clocks read a zero time the observers start moving in the positive direction of the OX axis with different proper accelerations $g_i$. Let $R'_i(g_i)$ be one of these observers. In accordance with the special relativity theory the acceleration $g_i$ can change in the range of values $0 < g_i < \infty$. This particular motion of observer $R'_i(g_i)$ relative to frame K (also known as the hyperbolic motion[2]) is described by the equation:

$$x = \frac{c^2}{g}\left(\sqrt{1 + \frac{g^2 t^2}{c^2}} - 1\right) \qquad (1)$$

whereas his instantaneous velocity relative to the same frame is given by

$$V_i = \frac{g_i t}{\sqrt{1 + \frac{g_i^2 t^2}{c^2}}}. \qquad (2)$$

**Figure 1** shows the world lines $WLR'_i(g_i)$ on a space-time diagram which displays in true values the space-time coordinates measured by observers of the K frame. Let $R_0$ be an observer at rest relative to the K frame and located at its origin O. His world line is $WLR_0$ and coincides with the time axis of the diagram. Here WLc represents the world line of a light signal emitted at *t=0* from the origin O in the positive direction of the OX axis. We present in Figure 1 also the asymptotes of this world line which are parallel to WLc.

The first problem we are confronted with is to find out the geometric locus of the points in the space-time diagram where all the velocities of the accelerating observers are equal to *V*, that is to say the iso-velocity lines. Eliminating $g_i$ between (1) and (2) we obtain that this locus is given by

$$\frac{x}{c} = \frac{1-\sqrt{1-\frac{V^2}{c^2}}}{\frac{V}{c}} t \qquad (3)$$

and we present some of them in Figure 1.

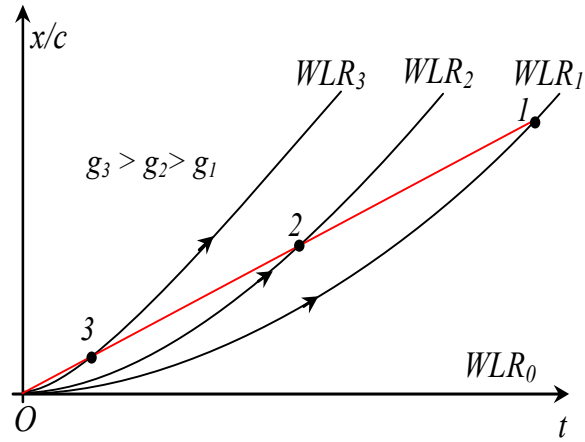

**Figure 1.** *The system of accelerating observers and an iso-velocity line*

Such a geometric locus intersects the world lines of the accelerating observers generating the events **1, 2** and **3** respectively. We use equation (2) in order to find out the space-time coordinates of the event associated with the fact that the velocity of an observer $R'(g)$ becomes equal to *V*. We obtain the time coordinate of this event as:

$$t = \frac{V}{g\sqrt{1-\frac{V^2}{c^2}}} \qquad (4)$$

and its space coordinate as:

$$x = \frac{c^2}{g}\left(\frac{1}{\sqrt{1-\frac{V^2}{c^2}}} - 1\right). \qquad (5)$$

Consider a very long platform that moves with constant velocity *V* in the positive direction of the OX axis and let K'(X'O'Y') be the inertial reference frame commoving with this platform. The origin O' of frame K' is located at the middle point of the platform. At the time *t* given by (4), when the accelerating observer attains the velocity *V,* it can smoothly land on the platform. Detected from the frame K' and in accordance with the Lorentz-Einstein transformations the space-time coordinates of the landing event are:

$$x' = \frac{c^2}{g}\left(1 - \frac{1}{\sqrt{1 - \frac{V^2}{c^2}}}\right) = -x \qquad (6)$$

$$t' = \frac{V}{g\sqrt{1 - \frac{V^2}{c^2}}} = t. \qquad (7)$$

**2. Differently accelerated twins without differential aging**

Some students still struggle with the idea that the acceleration is the root cause of differential aging in the classical twin paradox. To convince them about the contrary an ideal scenario should exhibit differential aging in the absence of differential acceleration between the twins. Some authors[3-14] have been successful in this regard but they ignore the small intervals of acceleration at the point of return or they need to alter the original scenario (one of the twins remains at home while the second travels away with high-velocity, subsequently turning around and speeding up to return home where it finds that its twin grew older).

A particular scenario known as the Boughn's version of the twins' paradox[10] maintains much of the original flavour of the classical paradox. In this experiment the twins experience identical acceleration and yet present differential aging. However, as pointed out by Desloge and Philpott[14] and later by Price and Gruber[12] the twins in the Boughn's scenario[10] perform a critical initial separation prior to their acceleration and they should be reunited at the end of their trip (to evaluate their ages) all of which complicates the possible causes of differential aging. Moreover, even in the Boughn's scenario some acceleration is needed to cause whatever aging is seen. Therefore, it still might appear to some students that the acceleration is the direct cause of different aging.

Attempting to find out the best way to convince someone that acceleration per se is not the cause of differential aging, Gruber and Price[13] analysed the opposite scenario: a simple twin paradox with the twins being subject to acceleration, but (in the limit) without exhibiting differential aging. This scenario involved a rocket undergoing periodic (harmonic) motion. The derived equations demonstrate that time dilation is dependent on the maximum velocity but not on acceleration. However one weak point of this scenario is that the harmonic motion can't be expressed by an easy to handle analytic function of the parameters characterizing the motion.

The accelerating observers $R'_i(g_i)$ that we introduced above land smoothly on the platform moving with velocity $V$ at the times

$$t_i = \frac{V}{g_i\sqrt{1-\frac{V^2}{c^2}}} \qquad (8)$$

Considering only two observers $R'_1(g_1)$ and $R'_2(g_2)$ ($g_2>g_1$) it is $R'_2(g_2)$ who lands first followed by $R'_1(g_1)$. After landing on the platform, the clocks attached to the accelerated observers start clicking in the rhythm of the synchronized clocks $C'_1$ and $C'_2$ of the frame K' that are located where the landings take place. The result is that at the instant when $R'_1(g_1)$ lands on platform both $C'_1$ and $C'_2$ display the same time and from now on they will display the same running time. It means that the two accelerating observers who started to move at the same time from the same point in space shall have the same age after landing.

We can now bring these two observers at the middle of the distance that separates their landing points, via some slow-speed transport, which doesn't affect their synchrony, thereby convincing them that even if they underwent different accelerations, they aged in the same way.

We have landed so far the observers who make up the system of uniformly accelerating observers onto a moving platform. We present now another scenario which enables us to move all observers from the origin O of the K frame where they are initially in a rest state to another point of the same reference frame where they achieve the rest state as well. This process involves an acceleration phase followed by a deceleration phase. Form classical physics we know the following problem: a particle starts moving from a state of rest performing an accelerating motion with constant acceleration $g$ in the positive direction of the OX axis of the inertial reference frame K(XOY). After a given time of motion $t_1$, when his velocity becomes $V$, the acceleration changes its sign instantaneously and after a second time of motion $t_2$ the particle stops. Let $L_1$ be the distance travelled during the period of acceleration and $L_2$ the distance travelled during the period of deceleration. We have

$$L_1 = \frac{gt_1^2}{2} \qquad (8)$$
$$V = gt_1 \qquad (9)$$

$$L_2 = Vt_2 - \frac{gt_2^2}{2} \tag{10}$$

where from we obtain

$$L_1 = L_2 \tag{11}$$

$$t_1 = \frac{V}{g} \tag{12}$$

and

$$t_2 = \frac{2V}{g} \tag{13}$$

resulting that the distances covered during the two motion phases and the time intervals during which they are performed are equal to each other.

Considering that the clock performs the hyperbolic motion, Iorio[15] studied the same problem and obtained for the physical quantities defined above the following results:

$$t_1 = \frac{V}{g\sqrt{1-\frac{V^2}{c^2}}} \tag{14}$$

$$L_1 = \frac{c^2}{g}\left(\frac{1}{\sqrt{1-\frac{V^2}{c^2}}} - 1\right) \tag{15}$$

$$t_2 = t_1 \tag{16}$$

and

$$L_2 = 2L_1. \tag{17}$$

where $V$ represents the maximal magnitude of the clock velocity during the scenario. The problem we are confronted with is to find out the reading of the clock after stopping. According to Iorio[15] we have

$$t'_{2,g} = \frac{2c}{g} a\tanh\frac{V}{c} = 2t'_{1,g} \tag{18}$$

where $t'_{1,g}$ represents the reading of the accelerating clock when its velocity is $V$. The important conclusion is that the time intervals given by (18) are linear proportional to $c/g$ and non-linear proportional to $V/c$.

Consider now the particular uniformly accelerating observers $R'_1(g)$ and $R'_2(2g)$, the second observer moving with twofold proper acceleration. During the time interval $t'_{2,2g}(2g)$ the observer $R'_2(2g)$ performs two successive cycles of acceleration-deceleration. Thereafter the clock he is caring reads the final time:

$$t'_{2,2g} = 2t'_{1,g} = \frac{2c}{g} a \tanh \frac{V}{c} \qquad (19)$$

i.e. the same time the clock of observer $R'_{1,g}$ reads when his landing takes place. The result is that we can simultaneously land at the same point in space all the uniformly accelerating observers $R'_i(2Ng)$ with $N$ integer and theirs clocks, who have started to move at $t=t'=0$ from the origin O, all aging in the same way.

The scenario we have followed is in accordance with the definition of the classical clock paradox and it shows that acceleration is not the main cause for the different aging of relativistic twins.

**3. Landing the observers of the uniformly accelerating reference frame**

The observers $R'_i(g_i)$ who make up the uniformly accelerating reference frame[14] perform the hyperbolic motion

$$x_i = \frac{c^2}{g_i} \sqrt{1 + \frac{g_i^2 t^2}{c^2}} \qquad (20)$$

starting at $t=0$ from the point located at a distance $c^2/g_i$ from the origin O of the K frame. The velocity of such an observer is given by (2). Eliminating $g_i$ between (2) and (19) we obtain that the geometric locus of the events associated with the fact that the velocity of the observer becomes equal to $V$ is given by

$$x = \frac{c^2}{V} t . \qquad (21)$$

The intersection of iso-velocity lines (21) with the world lines of the accelerating observers generates the events **1,2,3…** where the velocities of the corresponding observers is equal to $V$. Such an event is characterized by a time coordinate

$$t = \frac{V}{g\sqrt{1 - \frac{V^2}{c^2}}} \qquad (22)$$

and by a space coordinate

$$x = \frac{c^2}{g\sqrt{1-\frac{V^2}{c^2}}}.\qquad(23)$$

The landing of the observers on a platform that moves with constant velocity $V$ in the positive direction of the OX axis is smooth if they land whilst they velocity is exactly $V$. Let K'(X'O'Y') be the inertial reference frame attached to the platform. Detected from frame K' the space time coordinates of the landing events are

$$x'_i = \frac{c^2}{g_i} = x_i \qquad (24)$$

and

$$t'_i = 0. \qquad (25)$$

As we see from equation (24) the accelerating observers recover after landing the same longitudinal distribution they had initially along the X axis in the K frame, i.e. the landing of the observers who make up the uniformly accelerating reference frame doesn't change their positions in space. From equation (25) we also see that these observers didn't age differently. But because the observers didn't start from the same point in space this scenario doesn't meet the conditions imposed by the classical twin paradox.

### 5. Conclusions

The study of the landing of accelerating observers on a platform that moves with constant proper accelerations leads to interesting results concerning theirs ages and locations in the reference co-moving with the platform. Differently accelerating observers age in the same way in the conditions imposed by the classical definition of the clock paradox.